# Aberration-free three dimensional micromachining in glass with spatiotemporally shaped femtosecond laser pulses

**PENG WANG,**[1,2,3] **WEI CHU**[2,8] **WENBO LI,**[2,4] **YUANXIN TAN,**[2,3] **JIA QI,**[2,3] **YANG LIAO,**[2] **ZHANSHAN WANG,**[1] **YA CHENG**[2,5,6,7,9]

[1]*School of Physics Science and Engineering, Tongji University, Shanghai 200092, China.*

[2]*State Key Laboratory of High Field Laser Physics, Shanghai Institute of Optics and Fine Mechanics, Chinese Academy of Sciences, Shanghai 201800, China.*

[3]*University of Chinese Academy of Sciences, Beijing 100049, China.*

[4]*School of Physical Science and Technology, Shanghai Tech University, Shanghai 200031, China.*

[5]*State Key Laboratory of Precision Spectroscopy, School of Physics and Materials Science, East China Normal University, Shanghai 200062, China*

[6]*XXL - The Extreme Optoelectromechanics Laboratory, School of Physics and Materials Science, East China Normal University, Shanghai 200241, China*

[7]*Collaborative Innovation Center of Extreme Optics, Shanxi University, Taiyuan, Shanxi 030006, China*

[8]*Corresponding author: chuwei0818@qq.com*

[9]*Corresponding author: ya.cheng@siom.ac.cn*

**We observe that focusing a femtosecond laser beam simultaneously chirped in time and space domains in glass can efficiently suppress the optical aberration caused by the refractive index mismatch at the interface of air and the glass sample. We then demonstrate three dimensional microprocessing in glass with a nearly invariant spatial resolution for a large range of penetration depth between 250 μm and 9 mm without any aberration correction.**

## 1. INTRODUCTION

Femtosecond laser has become a powerful tool for three dimensional (3D) microfabrication [1-3]. One important application of femtosecond laser 3D micromachining is internal processing in transparent materials, i.e., glass and crystal, which has found great use in microfluidics and photonic integration [4, 5]. The longitudinal size of the microstructures that can be achieved with the femtosecond laser internal processing is usually limited by the working distance of the focal lens. By use of focal lenses of low numerical apertures (NAs), the working distance can be extended to enable 3D fabrication deeply into the transparent materials. Owing to the significant refractive index mismatch at the interface of air and the glass sample, the focal spot will be strongly distorted by the optical aberration with the increase of the penetration depth. An efficient solution on this issue is to use adaptive optics for correcting the deformation in the wavefront [6, 7]. Nevertheless, elongation of the focal spot along the optical axis of the focal lens intrinsically occurs at low NAs as a result of the diffraction, leading to degradation of the axial resolution which cannot be completely compensated for using the adaptive optics.

Recently, we have shown that the axial resolution at the low NAs can be substantially improved using simultaneously spatiotemporally focused (SSTF) femtosecond pulses [8]. The principle of SSTF has been well documented elsewhere [9-12]. Moreover, the aberration of SSTF-based two photon excitation with a high NA focal lens has been theoretically examined [13]. It should be noted that in the conventional SSTF scheme as frequently employed in the femtosecond laser micromachining so far, the optical layout is designed to ensure that the focused pulses are Fourier-transform-limited without any temporal chirp at the focus [14]. Here, we show that under the condition of focusing with a low-NA lens, one can maintain the high longitudinal resolution at any penetration depth within the glass sample up to the limit of the working distance of the focal lens by use of strongly chirped SSTF pulses. Remarkably, such an unusual characteristic is against the physical picture indicated by the simulation result obtained in the framework of linear propagation of the SSTF laser pulses. Therefore, a complete understanding on the observed phenomenon can only be achieved by developing a sophisticated three-dimensional simulation code taking into account of the nonlinear interaction between the laser pulses and glass as well as the nonlinear propagation effects, which is an attractive challenge for the theorists.

## 2. EXPERIMENT

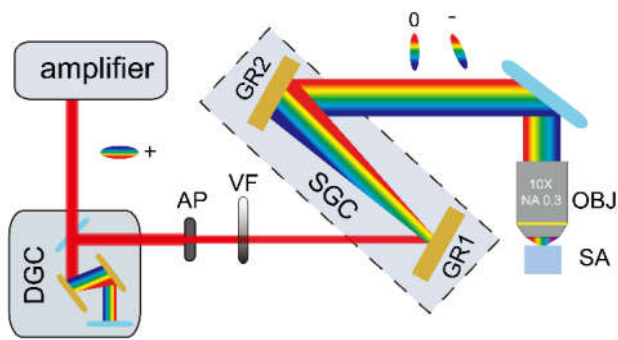

Fig. 1. Schematic of the experimental setup. DGC: double-pass grating compressor, AP: aperture, VF: variable neutral density filter, GR1~2: gratings, SGC: single-pass grating compressor, OBJ: objective lens, SA: sample.

The experimental setup is schematically illustrated in Figure 1. The laser pulses delivered by a commercial chirp pulse amplifier (CPA) system (Libra, Coherent, Inc.) were of a central wavelength of 800 nm and repetition rate of 1 kHz with a positive chirp. A double-pass grating compressor (DGC) was subsequently employed to adjust the temporal chirp of the femtosecond laser pulses. After passing through an aperture for trimming the beam, the femtosecond laser beam entered a single-pass grating compressor (SGC) which consists of two $\sigma$ = 830 grooves/mm gratings. The distance between the two gratings was fixed at ~ 210 mm, which could introduce a group delay dispersion (GDD) of -137555 fs$^2$ [15]. The total GDD of the pulse before the objective lens was controlled by synergetically tuning the DGC and SGC. The spatially dispersed laser pulses were then focused into the sample (i.e., a 10-mm-thick fused silica glass) using an objective lens (NA = 0.3). In the laser direct writing, the sample was translated with a resolution of 1 μm using a computer-controlled XYZ stage.

## 3. RESULTS AND DISCUSSION

Figure 2(a) presents the optical microscope image of the cross section of a modification track embedded 250 μm beneath the glass surface produced by linearly scanning a focal spot generated with the conventional focusing (CF) scheme, i.e., the femtosecond laser pulses were focused without the use of SGC. The average laser power was set at 1 mW so as to generate a peak intensity at the focus slightly above the threshold intensity of modification in the fused silica sample. All the modification tracks in Fig. 2 were written at a fixed scan speed of 50 μm/s. The result shows that with the CF scheme, the cross section of the modification track was highly asymmetric owing to an elongated focal spot [16].

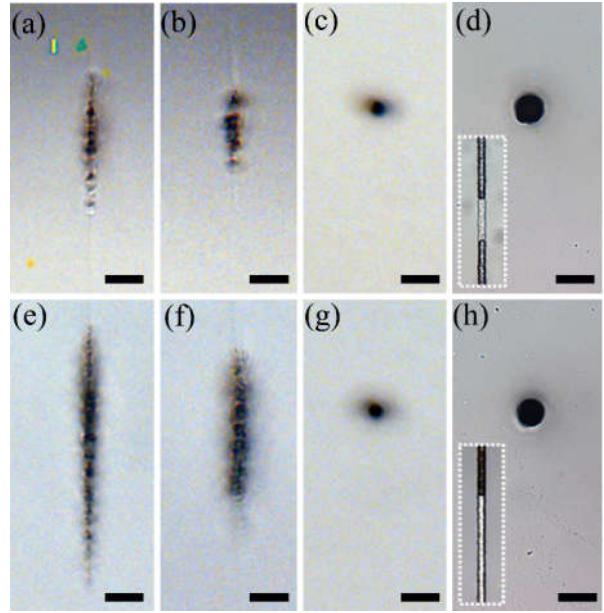

Fig. 2. The cross-sectional view optical micrographs of lines written with (a, e) CF, (b, f) TL-SSTF, and (c, g) TC-SSTF of a negative GDD. The penetration depths are 0.25 mm in (a, b, c), and 9 mm in (e, f, g). Shown in (d) and (h) are the optical images of the microchannels produced from (c) and (g), respectively. Insets in (d) and (h): top view micrographs of the hollow microchannels. Scale bar: 20 μm.

For comparison, the optical micrograph of the cross section of a modification track inscribed in the fused silica using the transform-limited SSTF (TL-SSTF) scheme was shown in Figure 2(b). In writing the structure in Fig. 2(b), the average laser power of the femtosecond laser was raised to 1.5 mW and the DGC was adjusted to completely compensate for the negative GDD induced by the SGC. The cross section of the modification track indicates a higher longitudinal resolution than that in Fig. 2(a) resulting from the characteristic of TL-SSTF. On the other hand, owing to the low NA of the objective lens, we found that it is difficult to obtain a completely spherical focal volume in glass despite of the great effort spent on the optimization of the focusing parameters. This is different from the previous result achieved with the TL-SSTF employing a high NA objective [8].

Remarkably, when we adjusted the DGC to introduce a GDD of -127260 fs$^2$ in the SSTF scheme, the cross section of the modification track appears to be nearly circular with a diameter of 10 μm as shown in Fig. 2(c). In this focusing condition, the laser power was raised to 4.4 mW (i.e., due to the chirp at the focus which leads to broadening in the pulse duration) to maintain the peak intensity slightly above the threshold intensity of modification in the fused silica. The dramatic improvement of the longitudinal resolution with the temporally chirped SSTF (TC-SSTF) scheme is further confirmed by the cross section of a hollow microchannel in Fig. 2(d), which is of a nearly perfect circular cross section as a direct indication of the balanced transverse and longitudinal resolutions. The 1-mm-long hollow microchannel was fabricated by immersing the structure in Fig. 2(c) in a 10 mol/L KOH solution to selectively remove the materials in the laser irradiated region. As shown in the inset, the black region of the microchannel is occupied by water left behind by the chemical wet etching.

For the low-NA focal lens chosen for our investigation, it offers the potential to enable fabrication of 3D microstructures deeply buried inside glass thanks to its long working distance. In this case, the influence from aberration should be carefully examined. Figure 2(e)-(g) show the optical micrographs of the cross sections of the modification tracks embedded 9 mm beneath the surface of the fused glass, which were produced with the CF, TL-SSTF, and TC-SSTF schemes at different average laser powers of 1.6 mW, 2 mW and 4.4 mW, respectively. The results show that for both the CF and TL-SSTF schemes, the optical aberration makes a strong effect which is manifested by the deteriorated longitudinal resolutions. In contrast, with the TC-SSTF scheme, the spatial profiles of the cross section in Fig. 2(g) and (h) remain nearly unchanged as compared to their respective counterparts in Fig. 2(c) and (d). Obviously, this feature is against the intuition.

To understand the physics behind the depth-independent direct writing achieved with the TC-SSTF scheme, we further examined the dependence of the cross section of the modification tracks on the GDD of the incident laser pulse. In this case, the microstructure was fabricated 4 mm beneath the glass surface (i.e., around the middle between the two penetration depths of 250 μm and 9 mm as shown above). The cross sectional optical micrographs of the modification tracks in Figure 3(a-e) were written at GDD = 0, -84840 fs$^2$, -106050 fs$^2$, -127260 fs$^2$, and -169680 fs$^2$, and average laser powers of 1.8 mW, 3.6 mW, 4 mW, 4.4 mW, and 5.2 mW, respectively. Under the conditions, the peak intensities were again slightly above the threshold intensities of the modification. As shown in Figure 3(a)-3(c), the increase of the GDD leads to the extension of the focal volume in the longitudinal direction. However, when the GDD increases to -127260 fs$^2$ and -169680 fs$^2$, a sudden change in the spatial profile was observed in Figure 3(d) and (e), in which a beautiful round-shaped cross section is recorded. Furthermore, for the GDD of -169680 fs$^2$, we raised the laser power to 7.4 mW. It was observed that two isolated modification zones separated by 55 μm along the longitudinal direction appear near the focus, as shown in Figure 3(f).

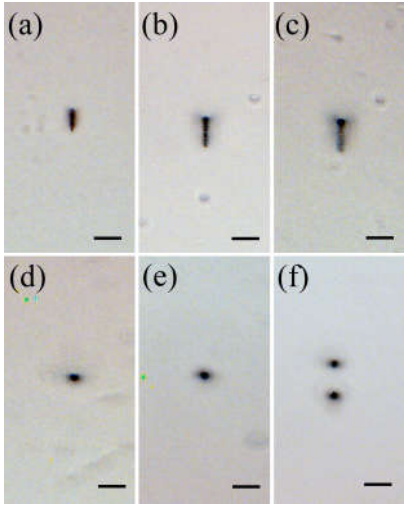

Fig. 3. The cross-sectional view optical micrographs of the lines written at different GDDs with SSTF scheme: (a) 0 fs$^2$, (b) -84840 fs$^2$, (c) -106050 fs$^2$, (d) -127260 fs$^2$, and (e, f) -169680 fs$^2$. The average power is 5.2 mW in (a-e) but it increases to 7.4 mW in (f). Scale bar: 50 μm.

The physics behind the observed behavior is yet to be clarified. Actually, owing to the strong couplings between the spatial and temporal coordinates of the SSTF light fields, three-dimensional simulation codes must be developed to enable investigation of *nonlinear* propagation of such laser beams. This is extremely challenging due to the huge demand on computational resources [17]. For this reason, we perform a simulation assuming the *linear* propagation of the SSTF pulses in glass, as we expect that the results may shed some insight on the mechanism behind the experimental observations in Fig. 2 and 3.

Using the analytical expressions in Ref. 18, the evolution of the SSTF pulse duration can be written as:

$$\tau_{bw}(\varepsilon) = \tau_0 \sqrt{\frac{1+\beta_{BA}^2\varepsilon^2}{1+\varepsilon^2}}, \quad (1)$$

where $\varepsilon = z/z_r$, z is the propagation position along axial direction, $z_r$ = 53 μm is the Rayleigh range, $\tau_0$ = 27 fs is the pulse duration of the original Fourier-transform limited Gaussian pulse, $\beta_{BA} = \sqrt{1+\beta^2}$, $\beta$ = 4 is the spatial chirp rate. When the initial input pulse is chirped before entering the focal lens, the on-axis duration of the propagating pulses can be derived as:

$$\tau(\varepsilon, \emptyset_{in}) = \tau_{bw}(\varepsilon)\sqrt{1 + (\Delta\omega^2 \frac{1+\varepsilon^2}{1+\beta_{BA}^2\varepsilon^2})^2(\emptyset_{in} - \frac{\beta^2\varepsilon}{\Delta\omega^2(1+\varepsilon^2)})^2}, \quad (2)$$

where $\Delta\omega = 7.5 \times 10^{13} \ rad/s$ is the $1/e^2$ half-width of the Gaussian input spectrum, and $\emptyset_{in}$ is the input second order phase of the chirped pulses (i.e., GDD). The spatial distributions of the peak intensities were calculated for three pulses with different initial GDDs of 0 fs$^2$, -84840 fs$^2$, and -127260 fs$^2$, as shown in Fig. 4(a)-(c), respectively. According to our calculation, with the TL-SSTF scheme, the Rayleigh length can be greatly reduced, as shown in Fig. 4(a). However, as shown in Fig. 4(b) and 4(c), with the TC-SSTF scheme, the focal spot appears highly elliptical, which is conspicuously different from the experimental observations. For comparison, the calculated on-axis temporal durations of the pulses under the respective focusing conditions were presented in Fig. 4(d)-(f). In the TL-SSTF scheme (i.e., Fig. 4(d)), the shortest pulse duration can only be transiently reached at the focus where the different frequency components overlap in space without any temporal chirp. In dramatic contrast, the on-axis duration of the focused pulse with the initial chirp of -84840 fs$^2$ undergoes a continuous temporal broadening during the propagation until reaching the maximum duration of ~13 ps near the focus. After the focus, the on-axis pulse duration decreases along the propagation direction until reaching the initial local bandwidth-limited duration of ~3 ps, as shown in Fig. 4(b). This feature is further highlighted by adding a larger initial GDD to the pulse as shown in Fig. 4(c), where the distribution of the peak intensity appears similar to that in Fig. 4(b) whereas a significantly longer pulse duration of ~20 ps has been achieved near the focus, as evidenced in Fig. 4(f). The results in Fig. 4 indicate the complexity of the physics behind the experimental results in Fig. 3, which should take into account of the nonlinear interactions between the femtosecond laser pulses and glass, such as the nonlinear Kerr effect, photoionization and avalanche ionization, plasma shielding and defocusing, *etc* [19].

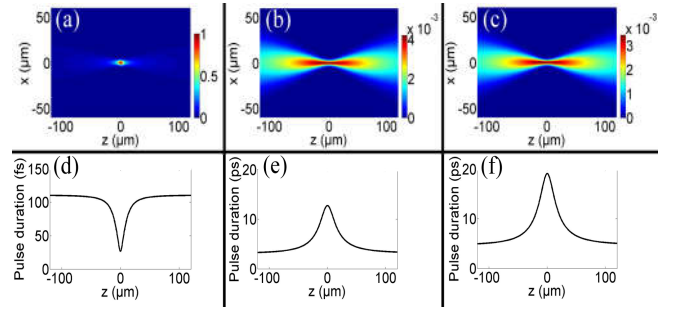

Fig. 4. Numerically calculated laser intensity distributions at the focus with different GDDs: (a) 0 fs$^2$, (b) -84840 fs$^2$, and (c) -127260 fs$^2$. (d), (e), (f) are the on-axis pulse durations in (a), (b), and (c), respectively.

## 4. CONCLUSION

To conclude, the spatiotemporal coupling of the ultrafast laser pulses provides a powerful approach to shaping the focal volume in ultrafast laser processing, which offers a unique characteristic of isotropic 3D fabrication resolution. It comes as a major surprise that the fabrication resolution remains nearly unchanged against the spherical aberration by adding an initial temporal chirp on the SSTF pulses. The observation suggests an incomplete understanding on the physics behind the interaction of spatiotemporally shaped laser pulses with transparent materials. From the application point of view, the result is particular useful to enable fabrication of 3D microstructures of large longitudinal sizes in glass, which are highly in demand for microfluidics, photonic integration, infrared and Terahertz optics, tissue engineering, and 3D glass printing, to name a few.

**Funding**. National Basic Research Program of China (2014CB921303), National Natural Science Foundation of China (11674340, 11734009, 61327902, 61590934, 61675220, 61505231), Strategic Priority Research Program of the Chinese Academy of Sciences (XDB16030300), Key Research Program of Frontier Sciences, Chinese Academy of Sciences (QYZDJ-SSW-SLH010), Project of Shanghai Committee of Science and Technology (17JC1400400) and Shanghai Rising-Star Program (17QA1404600).

**REFERENCES**

1. R. R. Gattass and E. Mazur, "Femtosecond laser micromachining in transparent materials," Nat. Photonics **2**, 219-225 (2008).
2. R. Osellame, H. J. W. M. Hoekstra, G. Cerullo, and M. Pollnau, "Femtosecond laser microstructuring: an enabling tool for optofluidic lab-on-chips," Laser Photon. Rev. **5**, 442-463 (2011).
3. K. Sugioka and Y. Cheng, "Femtosecond laser three-dimensional micro- and nanofabrication," Appl. Phys. Rev. **1**, 041303 (2014).
4. K. Itoh, W. Watanabe, S. Nolte, and C. B. Schaffer, "Ultrafast processes for bulk modification of transparent materials," MRS Bulletin **31**, 620-625 (2006).
5. K. Sugioka and Y. Cheng, "Femtosecond laser processing for optofluidic fabrication," Lab Chip **12**, 3576-3589 (2012).
6. B. P. Cumming, A. Jesacher, M. J. Booth, T. Wilson, and M. Gu, "Adaptive aberration compensation for three-dimensional micro-fabrication of photonic crystals in lithium niobate," Opt. Express **19**, 9419-9425 (2011).
7. A. Jesacher and M. J. Booth, "Parallel direct laser writing in three dimensions with spatially dependent aberration correction," Opt. Express **18**, 21090-21099 (2010).
8. F. He, H. Xu, Y. Cheng, J. Ni, H. Xiong, Z. Xu, K. Sugioka, and K. Midorikawa, "Fabrication of microfluidic channels with a circular cross section using spatiotemporally focused femtosecond laser pulses," Opt. Lett. **35**, 1106-1108 (2010).
9. D. N. Vitek, D. E. Adams, A. Johnson, P. S. Tsai, S. Backus, C. G. Durfee, D. Kleinfeld, and J. A. Squier, "Temporally focused femtosecond laser pulses for low numerical aperture micromachining through optically transparent materials," Opt. Express **18**, 18086-18094 (2010).
10. W. Chu, Y. Tan, P. Wang, J. Xu, W. Li, J. Qi, and Y. Cheng, "Centimeter-height 3D printing with femtosecond laser two-photon polymerization," https://doi.org/10.1002/admt.201700396.
11. F. He, B. Zeng, W. Chu, J. Ni, K. Sugioka, Y. Cheng, and C. G. Durfee, "Characterization and control of peak intensity distribution at the focus of a spatiotemporally focused femtosecond laser beam," Opt. Express **22**, 9734-9748 (2014).
12. C. Jing, Z. Wang, and Y. Cheng, "Characteristics and Applications of Spatiotemporally Focused Femtosecond Laser Pulses," Appl. Sci. **6**, 428 (2016).
13. B. S. Sun, P. S. Salter, and M. J. Booth, "Effects of aberrations in spatiotemporal focusing of ultrashort laser pulses," J. Opt. Soc. Am. A **31**, 765-772 (2014).
14. A. Patel, V. T. Tikhonchuk, J. Zhang, and P. G. Kazansky, "Non-paraxial polarization spatio-temporal coupling in ultrafast laser material processing," Laser Photon. Rev. **11**, 1600290 (2017).
15. S. Backus, C. G. Durfee, M. M. Murnane, and H. C. Kapteyn, "High power ultrafast lasers," Rev. Sci. Instrum. **69**, 1207-1223 (1998).
16. Y. Cheng, K. Sugioka, K. Midorikawa, M. Masuda, K. Toyoda, M. Kawachi, and K. Shihoyama, "Control of the cross-sectional shape of a hollow microchannel embedded in photostructurable glass by use of a femtosecond laser," Opt. Lett. **28**, 55-57 (2003).
17. V. P. Zhukov and N. M. Bulgakova, "Asymmetry of light absorption upon propagation of focused femtosecond laser pulses with spatiotemporal coupling through glass materials," Proc. SPIE **10228**, 102280D (2017).
18. C. G. Durfee, M. Greco, E. Block, D. Vitek, and J. A. Squier, "Intuitive analysis of space-time focusing with double-ABCD calculation," Opt. Express **20**, 14244-14259 (2012).
19. A. Couairon, and A. Mysyrowicz, "Femtosecond filamentation in transparent media", Phys. Rep. **441**, 47-189 (2007).